\documentclass[usenatbib]{mn2e}

\usepackage{amsmath, amssymb}
\usepackage[dvips]{graphicx}
\usepackage{epsfig}

\newcommand{\rlight}{r_{\rm L}}

\newcommand{\me}{m_{\rm e}}

\newcommand{\aap}{A\&A}
\newcommand{\mnras}{MNRAS}

\newcommand{\apj}{ApJ}
\newcommand{\aj}{Astr. J.}
\newcommand{\apjl}{ApJL}
\newcommand{\apjs}{ApJS}
\newcommand{\aplett}{Astrophysical Letters}

\title[High-energy emission from the pulsar striped wind]{High-energy
  emission from the pulsar striped wind: a synchrotron model for
  gamma-ray pulsars}

\author[J.  P\'etri]{J\'er\^ome P\'etri$^1$ \\
 $^1$Observatoire Astronomique de Strasbourg, UMR 7550 Universit\'e de Strasbourg, CNRS, 11 rue de l'Universit\'e, 67000 Strasbourg, France }

\begin{document}

\date{Accepted . Received ; in original form \today}

\maketitle

\begin{abstract}
  Gamma-ray pulsars constitute a class of high and very high-energy
  emitters for which the known population is steadily increasing
  thanks to the Fermi/Large Area Telescope. Nowadays, more than a
  hundred such pulsars have been detected, offering a reasonable
  sample onto which to apply statistical techniques in order to
  outline relevant trends in the averaged properties of this (maybe
  not so) special class of pulsars. In this paper, their gamma-ray
  luminosity and spectral features are explained in the framework of
  synchrotron radiation from particles located in the stripe of the
  pulsar wind. Apart from radiative losses, particles are also subject
  to a constant re-acceleration and reheating for instance by a
  magnetic reconnection induced electric field. The high-energy
  luminosity scales as $L_\gamma \approx 2\times10^{26} \textrm{ W} \,
  (L_{\rm sd}/10^{28} \textrm{ W})^{1/2} \, (P/1 \textrm{ s})^{-1/2}$
  where $L_{\rm sd}$ is the pulsar spindown luminosity and $P$ its
  period. From this relation, we derive important parameters of pulsar
  magnetosphere and wind theories.  Indeed, we find bulk Lorentz
  factor of the wind scaling as $\Gamma_{\rm v} \approx 10 \,
  \tau_{\rm rec}^{1/5} \, (L_{\rm sd}/10^{28} \textrm{ W})^{1/2}$,
  pair multiplicity~$\kappa$ related to the magnetization
  parameter~$\sigma$ by $\kappa\,\sigma \, \tau_{\rm rec}^{1/5}
  \approx 10^8$, and efficiency~$\eta$ of spin-down luminosity
  conversion into particle kinetic energy according to the relation
  $\eta\,\sigma\approx1$.  A good guess for the associated
  reconnection rate is then $\tau_{\rm rec} \approx 0.5 \, (L_{\rm
    sd}/10^{28} \textrm{ W})^{-5/12}$. Finally, pulses in gamma-rays
  are visible only if $L_{\rm sd}/P\gtrsim 10^{27} \textrm{
    W/s}$. This model differs from other high-energy emission
  mechanisms because it makes allowance not only for rotational
  kinetic energy release but also for an additional reservoir of
  energy anchored to the magnetic field of the stripe and released for
  instance by some magnetic reconnection processes.
\end{abstract}


\section{Introduction}
\label{sec:Intro}

Since their discovery in the radio band more than 40 years ago,
pulsars have now been firmly detected in all wavelengths of the
electromagnetic spectrum, from radio through optical, X-ray and
eventually high-energy gamma-rays. Recently the Crab was even detected
in very-high energies by VERITAS~\citep{2011Sci...334...69V} and
MAGIC~\citep{2011ApJ...742...43A}.  Although the radio emission
mechanism remains poorly understood, the MeV-GeV pulsed emission puts
severe constraints on the high-energy counterpart radiation
models. The new catalog of gamma-ray pulsars obtained by the Fermi-LAT
instrument \citep{2010ApJS..187..460A} increased the number of
gamma-ray pulsars from seven to about fifty. Since then, new pulsars
are discovered regularly, more than hundred are listed
nowadays~\citep{2012ApJS..199...31N}. This allows for the first time a
reasonable statistical analysis of the high-energy emission properties
of these objects like spectral shapes, cut-off energies, and
comparison between radio and gamma-ray radiation if both are
available.

It was long suspected that polar cap or outer/slot gaps could explain
this emission. Nevertheless, recent Fermi observations clearly
disfavored the polar cap explanation~\citep{2010ApJS..187..460A}. Some
of these gamma-ray pulsars do not significantly show a spectral
cut-off around a few GeV but rather a significant change in the
spectral index of the power law. Moreover MAGIC/VERITAS detection of
the Crab above 100 GeV seems to rule out outer gap models because the
accelerating electric field combined to radiation reaction limited
flow renders it difficult to observe photons above a few GeV. The
determination of the precise location of the emission regions is still
problematic. The in phase pulsation observed in both gamma-ray and in
radio for some millisecond pulsars \citep{2012ApJ...744...33G}
furnishes one more apparent contradiction between radio and
high-energy emission mechanisms. Do they eventually originate in the
same place in the magnetosphere or in the wind? But why are then some
gamma-ray pulsars radio quiet as those seen by
\cite{2012ApJ...744..105P}. We are still far from a consensus on the
emission model, be it in the radio band or in gamma-rays.

Radio pulses and gamma-ray photons are expected to be produced in
different emission sites, probably close to the neutron star surface
for the former, described by a polar cap model
\citep{1969ApL.....3..225R}, and in the vicinity of the light-cylinder
for the latter, explained by outer gaps \citep{1986ApJ...300..500C}.

Recently, gamma-ray light-curves have been computed for a realistic
magnetospheric model based on 3D MHD simulations of the near pulsar
magnetosphere \citep{2010ApJ...715.1282B}. In this model, gamma rays
are expected close to the light-cylinder.

An alternative site for the production of pulsed radiation has been
investigated a few years ago by~\cite{2002A&A...388L..29K}. This model
is based on the striped pulsar wind, originally introduced by
\cite{1990ApJ...349..538C}.  Emission from the striped wind originates
outside the light cylinder and relativistic beaming effects are
responsible for the phase coherence of this radiation
\citep{2009A&A...503...13P, 2011MNRAS.412.1870P}


In this paper, we show that the pulsed high-energy emission up to a
few GeV as observed by Fermi/LAT can be explained by synchrotron
radiation emanating from the relativistically hot current sheet
present in the pulsar striped wind. Our approach follows the study
done by \cite{1996A&A...311..172L} who included possible pair creation
which are discarded in the present work. Sec.~\ref{sec:Modele}
describes the dynamics and geometry of the relativistic plasma flow as
well as the radiation model. Then Sec.~\ref{sec:Resultat} derives some
important constraints on the bulk Lorentz factor of the wind, the pair
multiplicity factor, the magnetization parameter as well as a
criterion for gamma-ray pulsar detection. Conclusions are drawn in
Sec.~\ref{sec:Conclusion}.

\section{The synchrotron striped wind model}
\label{sec:Modele}

Synchrotron emission from the striped wind applies successfully to the
pulsed optical polarization properties of the Crab
pulsar~\citep{2005ApJ...627L..37P}. Our aim is to extend to the
highest possible energies the synchro-photons emanating from the
current sheet. In the present work, we are not concerned with the
phase-resolved emission nor in the polarization properties, but we
only focus on the average gamma-ray features of the Fermi/LAT detected
pulsars depending on two fundamental observables: the period~$P$ of
the pulsar and its first derivative~$\dot{P}$. We emphasize that this
two-parameter family of model remains very restrictive and will not be
able to explain in detail each individual pulsar. Indeed, the
mechanisms occurring in the current sheet are far from being
understood, as they involve micro-physics that has not yet been
addressed self-consistently and linked to the overall structure and
dynamics of the magnetosphere and wind. Nevertheless, as we will show,
our model is able to reproduce qualitatively and quantitatively the
full set of Fermi data with reasonable accuracy.  The derived
parameters such as wind Lorentz factor, pair multiplicity and
magnetization are consistent with values obtained from other
independent considerations, giving us confidence in our description of
the emission mechanism.

\subsection{Plasma configuration}

At the heart of any pulsar emission model, be it radio, optical or
X-rays/gamma-rays, there is a (possibly relativistic) plasma flowing
in a strongly magnetized field, preferentially within the
magnetosphere for the polar cap and slot/outer gaps. Contrary to these
models, we assume that the pulsed emission is produced within the
current sheet of the striped wind, thus well outside the
magnetosphere, a place where the magnetic configuration switches from
poloidal dominant to toroidal dominant. The plasma in this current
sheet is the crucial ingredient in our model. It is embedded in a
mostly non emitting cold plasma. More precisely, the striped wind flow
is made of two distinct parts, namely
\begin{itemize}
\item a cold and strongly magnetized plasma of particle density number
  $n_{\rm c}$ and Lorentz factor $\Gamma_{\rm v}$, as measured in the
  lab frame.  Note that $\Gamma_{\rm v}$ corresponds to the bulk
  Lorentz factor of the whole striped wind structure. This part of the
  wind does not radiate significantly.
\item a hot and tenuous but weakly magnetized plasma of particle
  density number $n_{\rm h}$ occupying a fraction $\Delta$ of the
  wavelength of the wind $\lambda_{\rm v} = 2\,\pi\,\beta_{\rm
    v}\,\rlight$, also measured in the lab frame. The fraction of hot
  plasma is negligible with respect to the cold part, i.e.~$\Delta \ll
  1$. $\rlight=c/\Omega$ is the light-cylinder radius, $\beta_{\rm
    v}=v/c$ the wind speed normalized to the speed of light~$c$ and
  $\Omega$ the stellar rotation rate.
\end{itemize}
Later on, it will be useful to deal with the proper densities and
proper lengths given by $n_{\rm c} = \Gamma_{\rm v} \, n'_{\rm c}$,
$n_{\rm h} = \Gamma_{\rm v} \, n'_{\rm h}$ and $\lambda'_{\rm v} =
\Gamma_{\rm v} \, \lambda_{\rm v}$. Proper quantities in the rest
frame of the wind are always primed except for thermodynamical
quantities such as pressure and temperature which are meaningful only
in the proper frame. Furthermore, we assume that pairs are created
within the magnetosphere and cool down quickly before reaching the
wind zone. Therefore, they exclusively replenish the cold part of the
striped wind. Acceleration and heating of particles occurs once in the
stripe and not before. This is clearly an additional source of energy,
a kind of magnetic luminosity, independent of the spin-down luminosity
of the pulsar. Therefore, rotational kinetic energy is spent to
produce relativistic leptons with a curvature radiation reaction
limited Lorentz factor of about $\gamma\approx10^7$. They will cool
down in the magnetized part of the wind because of magnetic field
strength around the light-cylinder higher than $10^{-2} \textrm{ T}$
according to Fermi/LAT first pulsar catalog
\citep{2010ApJS..187..460A}.  Indeed, the synchrotron cooling time
$\tau_{\rm syn} \approx 7.7 \textrm{ s} \, \gamma^{-1} \, \left(
  \frac{B}{1\textrm{ T}} \right)^{-2}$ is much less than the period of
a pulsar.  In the most unfavored case, radiation losses are still such
that $\tau_{\rm syn} \lesssim 1 \textrm{ ms}$. Thus, our model can
apply irrespective of the period of the pulsar, millisecond or normal.

\subsection{Wind flow}

Although the hot plasma is much less populated than the cool part, it
radiates significantly more because of its relativistic
temperature. Indeed, to ensure a quasi-stationary equilibrium state,
magnetic pressure outside the stripe has to be compensated by gaseous
pressure inside the current sheet. Thus in the wind frame, pressure
balance implies
\begin{equation}
  \label{eq:Equilibre_Pression}
  \frac{1}{3} \, \gamma_{\rm h}' \, n'_{\rm h} \, \me \, c^2 = \frac{B'^2}{2\,\mu_0}
\end{equation}
$\me$ is the electron rest mass and $\mu_0$ vacuum permeability.  We
recall that the magnetic field is essentially toroidal in the wind
zone, a Lorentz boost thus gives $B'=B/\Gamma_{\rm v}$ where
$B=B_L\,\rlight/r$, $B_L$ being the magnetic field strength at the
light-cylinder. The Lorentz factor $\gamma_{\rm h}'$ can be deduced
from considerations about the dynamics in the current sheet as
explained in the next paragraph (see Eq.(\ref{eq:gamma_max})), leading
to the knowledge of the density of the hot plasma~$n_{\rm h}'$ as
given by the pressure balance Eq.~(\ref{eq:Equilibre_Pression}).

The bulk kinetic energy of the wind is extracted from the spin-down
luminosity of the pulsar. Noting~$\eta$ the efficiency coefficient of
rotational to particle energy conversion, the plasma kinetic energy
density is
\begin{equation}
  \label{eq:Gamma_v_Lsd}
  \Gamma_{\rm v} \, n_{\rm c} \, \me\, c^2 = \eta \, \frac{L_{\rm sd}}{4\,\pi\,r^2\,c}
\end{equation}
Note that we neglect the contribution of the hot component because
$\Delta \ll 1$ and $L_{\rm sd} = 4\,\pi^2\,I\,\dot{P}\,P^{-3}$ is the
spin-down luminosity and $I=10^{38} \textrm{ kg\,m}^2$ the neutron
star moment of inertia.  Moreover, the cold plasma density is related
to the pair creation rate~$\dot{N}_\pm$ (particles generated per
second) by $4\,\pi\,r^2\,c\,n_{\rm c} = \dot{N}_\pm$ where the
contribution from both poles are taken into account and
\begin{equation}
\label{eq:Ndot}
  \dot{N}_\pm \approx 2.77 \times 10^{30} \textrm{ s}^{-1} \, \kappa \, \left(
    \frac{P}{1~s} \right)^{-2} \, \left( \frac{B_{\rm ns}}{10^8\text{ T}}
  \right) \, \left( \frac{R_{\rm ns}}{10\text{ km}} \right)^3
\end{equation}
$\kappa$ is the usual pair multiplicity factor, $e$ the absolute value
of the charge of an electron, $\varepsilon_0$ the vacuum permittivity,
$B_{\rm ns}$ the stellar surface magnetic field intensity and $R_{\rm
  ns}$ the neutron star radius. Inserting Eq.~(\ref{eq:Ndot}) into
Eq.~(\ref{eq:Gamma_v_Lsd}), we found a relation between $\Gamma_{\rm
  v},\kappa$ and~$\eta$ such that
\begin{equation}
  \label{eq:Gamma_v_kappa_eta}
  \Gamma_{\rm v} \, \kappa \approx 8.7 \times 10^{8} \, \eta \,  \left( \frac{L_{\rm sd}}{10^{28} \textrm{ W}} \right)^{1/2}
\end{equation}
A similar relation holds for the magnetization defined by $\sigma =
B^2/(\mu_0\,\Gamma_{\rm v}\,n_{\rm c}\,\me\,c^2)$. The magnetization
parameter is approximately the inverse of the efficiency coefficient,
i.e.
\begin{equation}
  \label{eq:SigmaEta}
  ( 1 + \sin^2\chi ) \, \sigma \, \eta = 1
\end{equation}
where $\chi$ is the pulsar obliquity and assuming the force-free
spin-down luminosity of \cite{2006ApJ...648L..51S} and confirmed by
\cite{2012arXiv1205.0889P}. This reduces the number of independent
parameters in the model.  As expected, the fraction of spin-down
luminosity converted into particles remains weak $\eta\ll1$ as long as
the magnetization parameter remains large $\sigma\gg1$.

\subsection{Dynamics in the current sheet}

It is usually claimed that the total gamma-ray luminosity should not
exceed the spin-down luminosity. Actually, this assertion only holds if
the rotation of the neutron star is the unique source of
energy. Another non negligible reservoir could be the magnetic field
itself. We will indeed assume that part of the radiation is due to
magnetic reconnection within the current sheet. This will alleviate
the restriction $L_\gamma \leq L_{\rm sd}$ although the model not
necessarily violates this condition. It will be checked a posteriori.

Particles in the sheet lose energy due to synchrotron cooling.
However, a stationary state in the wind frame can be reached if
particles re-energize by acceleration in a reconnecting magnetic field
configuration. The key parameter is the collisionless reconnection
rate, $\tau_{\rm rec}$. It will be constrained from spectral features
of the Fermi/LAT data. This means that the induced electric field
re-accelerating particles is of the order $E_{\rm rec} \approx
\tau_{\rm rec}\,c\,B$.  Equaling the synchrotron cooling timescale
with the acceleration timescale in the reconnection layer, we find
that the maximum Lorentz factor of the leptons (we use primed quantity
to express the fact that they are evaluated in the wind frame) is
\begin{equation}
  \label{eq:gamma_max}
  \gamma_{\rm h}' = \sqrt{
    \frac{3}{2}\,\frac{\mu_0\,e\,c}{\sigma_T\,B_L'} \,
    \frac{r}{\rlight} \, \tau_{\rm rec} }
\end{equation}
Basically, it depends on the magnetic field strength at the light
cylinder and on the distance to the centre of the pulsar.  For these
Lorentz factors, the typical energy of a synchrotron photons as seen
in the wind frame is
\begin{equation}
  \label{eq:energ_syn_1}
  \varepsilon_B' = \frac{3}{2} \, \gamma_{\rm h}'^2 \, \frac{B'}{B_q}
  \, \me \, c^2 = \frac{9}{4} \, \frac{\mu_0\,e\,\me\,c^3}{\sigma_T\,B_q}
  \, \tau_{\rm rec}
\end{equation}
where $B_q \approx 4.4 \times 10^9 \textrm{ T}$ is the quantum
magnetic field.  Thus the synchro-photon energy does only depend
linearly on the reconnection rate. Interestingly the photon lies in
the MeV band because $\varepsilon_B' = 236 \textrm{ MeV} \, \tau_{\rm
  rec}$.  Back to the lab frame, the photon is boosted by a Lorentz
factor at most $2\,\Gamma_{\rm v}$ thus
  \begin{equation}
    \label{eq:Cutoff}
    \varepsilon_B = 2 \, \Gamma_{\rm v} \, \varepsilon_B' = 472 \textrm{ MeV} \, \Gamma_{\rm v} \, \tau_{\rm rec}.
  \end{equation}
  According to the Fermi/LAT catalog \citep{2010ApJS..187..460A}, the
  cut-off energy lies in the range 0.2-8~GeV. In our model, we
  interpret this cut-off more properly as a break in the spectral
  energy distribution. To make the idea more precise, think about an
  inverse Compton spectrum evolving from the Thomson regime to the
  Klein-Nishina regime. There is a break in the power law entailing
  precious information on the underlying emitting plasma. Here, the
  break is a consequence of the rapid cooling of electrons more
  energetic than $\gamma_{\rm h}'\,\me\,c^2$. Using a fiducial value
  of $\tau_{\rm rec} \approx 0.1$, we see that the Lorentz factor of
  the wind should be around $\Gamma_{\rm v} \in[5,10^2]$. We emphasize
  that the estimates given here should be understood as orders of
  magnitude roughly within a factor 10. More confidently we expect
  $\Gamma_{\rm v} \in [1,10^3]$ as will be shown from the
  observations. Moreover, the reliability of the value of the cut-off
  energy is probably doubtful at least for the most energetic pulsars
  like the Crab if not for almost the whole sample of gamma-ray
  pulsars.

\subsection{Synchrotron luminosity}

In the frame of the current sheet, relativistic particles in the hot
component radiate synchrotron power at a rate $P'_{\rm
  syn}(\gamma_{\rm h}',B') = (4/3) \, \sigma_T \, c \, \gamma_{\rm
  h}'^2 \, U_B'$ where $U_B'=B'^2/(2\,\mu_0)$ is the magnetic energy
density in the wind frame.  The total emissivity of the hot plasma,
assuming isotropy is $j'_{\rm syn}(\nu') = (n_{\rm h}'/(4\,\pi)) \,
P'_{\rm syn}(\gamma_{\rm h}',B') \, \delta(\nu'-\nu'_{\rm syn})$ where
$\nu'_{\rm syn} = (3/2)\,\gamma_{\rm h}'^2\,e\,B'/\me$.  By Lorentz
transformation and using relativistic invariants, the total
synchrotron luminosity,
boosted to the observer frame (expected to be interpreted as the
gamma-ray luminosity) is
\begin{equation}
  \label{eq:L_syn}
  L_{\rm syn} = 8.7 \times 10^{-20} \textrm{ W} \,
  \sqrt{\frac{\tau_{\rm rec}}{\alpha}} \, \frac{\Delta}{\Gamma_{\rm
      v}^{5/2}} \, L_{\rm sd}^{5/4} \, \sqrt{\frac{L_{\rm sd}}{P}}
\end{equation}
where $\Gamma_{\rm v}\geq1$ is the bulk Lorentz factor of the wind,
$\Delta\leq1$ the thickness of the current sheet in fraction of the
wavelength~$\lambda_{\rm v}$, and $\alpha \geq 1$ sets the minimum
distance where emission starts (in the region $r\ge\alpha\,\rlight$).
From eq.~(\ref{eq:Gamma_v_kappa_eta}), we get
\begin{equation}
  \label{eq:L_syn2}
  L_{\rm syn} = 3.9 \times 10^{-7} \textrm{ W} \,
  \sqrt{\frac{\tau_{\rm rec}}{\alpha}} \, \left( \frac{\kappa}{\eta} \right)^{5/2} \, \Delta \, \sqrt{\frac{L_{\rm sd}}{P}}
\end{equation}
Knowing the synchrotron luminosity Eq.~(\ref{eq:L_syn2}) and the
relations~(\ref{eq:Gamma_v_kappa_eta}) and~(\ref{eq:SigmaEta}), we are
able to deduce important constraints on the fundamental parameters.
Theoretically, for fixed $\{\alpha, \Delta, \kappa, \eta, \tau_{\rm
  rec}\}$, our model predicts $L_{\rm syn} \propto \sqrt{L_{\rm
    sd}/P}$.

\section{Results}
\label{sec:Resultat}

In this paragraph, we discuss the implication of the striped
synchrotron model and the latest Fermi/LAT detections and give
explicit values for the most relevant parameters of pulsar
magnetosphere and wind theories.

\subsection{Luminosity and spectral features}

\begin{figure}
  \centering
  \includegraphics[width=0.45\textwidth]{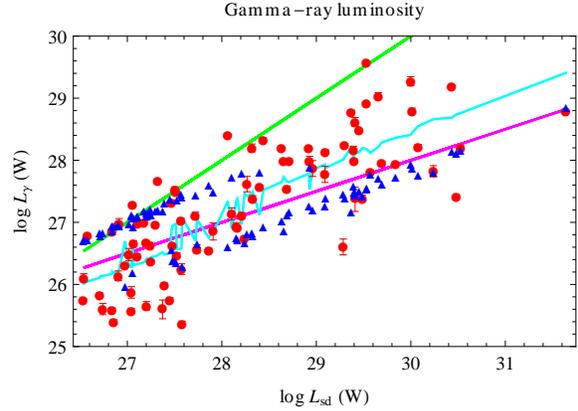}
  \caption{Pulsar gamma-ray luminosity~$L_\gamma$ versus spin-down
    luminosity~$L_{\rm sd}$. Fermi observations are depicted by red
    points whereas the predicted luminosity is presented by blue
    triangles. The magenta line corresponds to the law $L_\gamma =
    10^{26} \textrm{ W} \, ( L_{\rm sd} / 10^{26} \textrm{ W})^{1/2}$,
    the green line to $L_\gamma = L_{\rm sd}$ and the cyan line to the
    best fit Eq.~(\ref{eq:Fit}).}
  \label{fig:Luminosite}
\end{figure}

It is possible to relate $\Gamma_{\rm v}$ to the pair
multiplicity~$\kappa$ and to the magnetization~$\sigma$.  According to
the second source catalog~(2FGL) \citep{2012ApJS..199...31N}, a good
fit of the whole sample of gamma-ray pulsars following the formal
dependence as suggested by Eq.~(\ref{eq:L_syn2}) thus with respect to
$\sqrt{L_{\rm sd}/P}$ is
\begin{equation}
  \label{eq:Fit}
  L_\gamma \approx 2 \times 10^{26} \textrm{ W} \, \left( \frac{L_{\rm sd}}{10^{28} \textrm{ W}} \right)^{1/2} \, \left( \frac{P}{1\textrm{ s}} \right)^{-1/2}
\end{equation}
The constant of proportionality comes from Fermi/LAT data and will be
compared to the factor in front of $\sqrt{L_{\rm sd}/P}$ in
Eq.~(\ref{eq:L_syn2}). Note that the integral energy flux above
100~MeV from 2FGL and distances from ATNF are subject to several
sources of uncertainties
\citep{2005AJ....129.1993M}\footnote{see also the website from ATNF \\
  \textrm{http://www.atnf.csiro.au/research/pulsar/psrcat/expert.html}}. This
reflects into the derived quantities such as $\{\Gamma_{\rm v},
\tau_{\rm rec}, \Delta,\alpha\}$ for which the same remarks hold. We
also used a beaming factor of~1.  Thanks to the above fit,
Eq.~(\ref{eq:Fit}), and to the relation (\ref{eq:L_syn2}), we get $(
\kappa/\eta)^{5/2} \, \Delta \, \sqrt{\tau_{\rm rec}/\alpha} \approx 5
\times 10^{18}$.  Adopting typical values of $\Delta\approx0.1$,
$\alpha\approx 10$, the relation between pair multiplicity and
efficiency becomes $\kappa \, \tau_{\rm rec}^{1/5} \approx 1.2 \times
10^8 \,\eta $.  From this, we immediately deduce the bulk Lorentz
factor of the wind
\begin{equation}
  \label{eq:Gamma_v}
  \Gamma_{\rm v} \approx 10 \, \tau_{\rm rec}^{1/5} \, \left( \frac{L_{\rm sd}}{10^{28} \textrm{ W}} \right)^{1/2}
\end{equation}
It is proportional to the square root of the spin-down luminosity.  By
comparison between the cut-off energies observed by Fermi/LAT and the
prediction Eq.~(\ref{eq:Cutoff}), we can constrain the reconnection
rate for each pulsar such that
\begin{equation}
  \label{eq:tau_rec}
  \tau_{\rm rec} \approx \left( \frac{4.72 \textrm{ GeV}}{E_{\rm cut} (\textrm{GeV})}  \right)^{-5/6} \, \left( \frac{L_{\rm sd}}{10^{28} \textrm{ W}} \right)^{-5/12}
\end{equation}
The reconnection rate is about unity for low $L_{\rm sd}$ pulsars such
as the millisecond ones. This might be possible if the background pair
density is about~1\% of the current sheet particle density number
\citep{2012ApJ...750..129B}. We suspect magnetic dissipation to reach
its maximal possible value for those pulsars with $L_{\rm sd} \leq
10^{28} \textrm{ W}$ whereas $\tau_{\rm rec}$ decreases monotonically
down to~0.01 for the brightest pulsars with $L_{\rm sd} \geq 10^{28}
\textrm{ W}$. This splits the gamma-ray pulsars population into two
groups, a first with reconnection pushed to its highest limit and a
second with less drastic reconnection rate. Moreover, the photon
spectral indexes founded by Fermi/LAT fall in the range
$\Gamma\in[0.6,2]$. This is consistent with a synchrotron spectrum
emanating from a power law distribution of leptons with $p\in[1,3]$
and $dn_{\rm h}/d\gamma(\gamma) \propto \gamma^{-p}$. This could be
explained by a relativistic kinetic reconnection process, either very
localized close to the X-point where it is found that $p \approx 1$,
or on a global scale, within an extended region such as a simulation
box for PIC simulations \citep{2004ApJ...605L...9J} where
$p\approx3$. A detailed analysis of the kinetic reconnection processes
is out of the scope of this letter but see
\cite{2007ApJ...670..702Z}. Detailed knowledge of the particle
distribution functions within the current sheet awaits extensive PIC
simulations with radiation effects included
\citep{2009PhRvL.103g5002J}. The results of our model concerning the
expectation about the gamma-ray luminosity are summarized in the plot
Fig.~\ref{fig:Luminosite}. We show the gamma-ray luminosity versus the
spin-down luminosity for a large sample of gamma-ray pulsars. The
predicted luminosity, blue triangles, is overlapped to the Fermi/LAT
data, red points.  Although not explicitly imposed by our model and
actually not required, the gamma-ray luminosity nevertheless satisfies
$L_\gamma \lesssim L_{\rm sd}$.

\subsection{Condition for pulsed gamma-ray emission}

\begin{figure}
  \centering
  \includegraphics[width=0.4\textwidth]{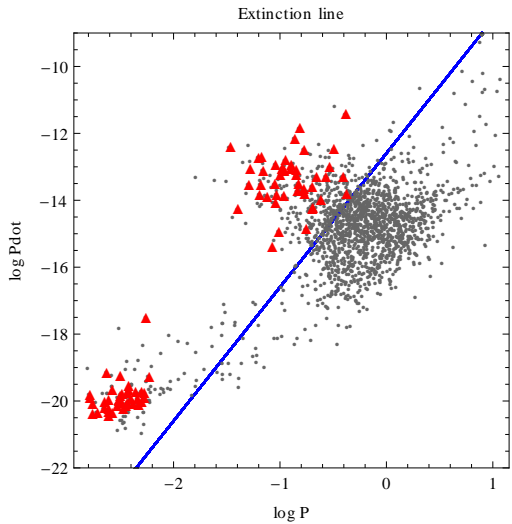}
  \caption{The extinction line for gamma-ray pulsation in the
    $P\dot{P}$~diagram. All radio pulsars are shown in grey points
    whereas Fermi/LAT detected gamma-ray pulsars are in red
    triangles. They all satisfy our criterion
    Eq.~(\ref{eq:CriterePulse}), lying above the blue extinction
    line.}
  \label{fig:Critere_Pulse}
\end{figure}

Gamma-ray emitters are identified as pulsars because of their very
stable clock. In our model, pulsation is a consequence of the
combination of relativistic beaming and striped wind
structure. Therefore, we expect to observe MeV-GeV pulsed emission
only if the stripe survives the destructive influence of magnetic
dissipation due to reconnection. Assuming as a first approximation
that the stripe resembles a relativistic Harris current sheet, a
stationary equilibrium can exist only if the drift speed of the
leptons within the sheet is less than the speed of light. This is
actually a criterion for reconnection as explained in
\cite{2003ApJ...591..366K}.  In the wind frame, the drift speed is
given by $\beta'_{\rm d} = r_B'/(\Delta\,\lambda_{\rm v}')$ where
$r_B'$ is the proper Larmor radius.  Replacing the bulk Lorentz factor
of the wind by the approximate relation Eq.~(\ref{eq:Gamma_v}), we
derive a simple criterion for pulsed emission from the condition
$\beta'_{\rm d}\le1$. According to its spin-down luminosity and to its
period, a pulsar will be detected as gamma-ray emitter if $L_{\rm
  sd}/P \gtrsim (10^{23} \textrm{ W/s} ) \, \tau_{\rm rec} \, \alpha^3 /
\Delta^2$.  For typical pulsar parameters, we found
\begin{equation}
  \label{eq:CriterePulse}
  \frac{L_{\rm sd}}{P} \gtrsim 10^{27} \textrm{ W/s}
\end{equation}
This criterion is shown in Fig.~\ref{fig:Critere_Pulse} for the
Fermi/LAT pulsar second catalog and marked as a blue line. All
detected pulsars fall into this criterion. The two populations of
pulsars, millisecond and normal, are clearly distinguishable. The gap
between the extinction line and the detected millisecond pulsar is due
to the $P^{-1}$ law in Eq.~(\ref{eq:CriterePulse}). The average
parameters $(\alpha,\Delta,\tau_{\rm rec})$ may also be slightly
different of those from normal pulsars. The absence of pulsation does
not imply that the underlying neutron star is not a gamma-ray emitter,
it just says that it will not be seen as a gamma-ray pulsar.

\section{Conclusion}
\label{sec:Conclusion}

In this paper, we showed that the pulsed MeV-GeV emission from
gamma-ray pulsars can be explained by a synchrotron model in the
striped pulsar wind including some magnetic reconnection. Relevant
average parameters of our model are given by $\Delta \approx 0.1$ and
$\alpha \approx 10$. The reconnection rate~$\tau_{\rm rec}$ and bulk
Lorentz factor of the wind~$\Gamma_{\rm v}$ are explicit functions
of~$L_{\rm sd}$. Moreover the predicted extinction line is in
agreement with gamma-ray pulsar observations.

Our synchrotron model predicts a clear break in the spectra around a
few GeV depending on the bulk Lorentz factor and reconnection rate. In
order to explain recent observations by VERITAS and MAGIC, this
picture should be supplemented with an inverse Compton
counterpart. The model would also benefit from phase-resolved
high-energy polarization measurements which are highly discriminating
for current models.

\section*{Acknowledgments}

I am grateful to David A. Smith for stimulating discussions and careful
reading of the manuscript.


\end{document}